\documentclass[11pt,eqs]{article}
\usepackage{latexsym}

\makeatletter
\newcommand\xleftrightarrow[2][]{%
  \ext@arrow 9999{\longleftrightarrowfill@}{#1}{#2}}
\newcommand\longleftrightarrowfill@{%
  \arrowfill@\leftarrow\relbar\rightarrow}
\makeatother

\title{
Advantage of the second-order formalism in double space T-dualization of type II superstring
\thanks{Work supported in part by
the Serbian Ministry of Education, Science and Technological Development, under contract No. 171031.}}
\author{B. Nikoli\'c and B. Sazdovi\'c
\thanks{e-mail: bnikolic, sazdovic@ipb.ac.rs}\\
{\it Institute of Physics,}\\
{\it University of Belgrade,}\\
{\it 11001 Belgrade, P.O.Box 57, Serbia}}

\begin{document}

\maketitle
\begin{abstract}
In this article we present bosonic T-dualization in double space of the type II superstring theory in pure spinor formulation. 
We use the action with constant background fields obtained from the general case under some physically and mathematically 
justified assumptions. Unlike Ref.\cite{bnbstip2}, where we used the first order theory, in this article fermionic momenta are 
integrated out. Full T-dualization in double space is represented as permutation of initial $x^\mu$ and T-dual coordinates $y_\mu$. 
Demanding that T-dual transformation law of the T-dual double coordinate ${}^\star Z^M=(y_\mu,x^\mu)$ is of the same form as for initial one $Z^M=(x^\mu,y_\mu)$,  
we obtain the form of the T-dual background fields in terms of the initial ones. Advantage of using the action with integrated fermionic momenta is 
that it gives all T-dual background fields in terms of the initial ones. In the case of the first order theory 
\cite{bnbstip2} T-dual R-R field strength was obtained out of double space formalism under additional assumptions.
\end{abstract}

\section{Introduction}
\setcounter{equation}{0}

T-duality is a feature which cannot be met in the point-particle theory and represents novelty brought by string theory \cite{S,B,RV,GPR,AABL}. The basic
mathematical framework, in which T-dualization is performed, is Buscher procedure \cite{B}.
The starting point of the procedure is existence of global isometries along some
directions. In the next step we localize that symmetry introducing world-sheet covariant derivatives (instead ordinary ones) and gauge fields.
In order to make gauge field to be unphysical degree of freedom, a term with Lagrange multipliers is added to the action.
The final phase of the procedure is using of gauge freedom to fix initial coordinates.
Variation of the gauge fixed action with respect to the Lagrange
multipliers produces the initial action, while variation with respect to the gauge fields gives T-dual action. Combining these equations of motion
the relations connecting initial and T-dual coordinates are obtained. These relations are known in literature as T-dual transformation laws.

Why T-duality is so important? The answer is in relation with understanding the M-theory. Five consistent superstring theories are connected by web of T and S-dualities. 
It is known fact in the case of type II superstring theories that T-dualization along one spatial dimension transforms type IIA(B) to type IIB(A) theory, 
while T-dualization along time-like direction produces type II${}^\star$ theory, which R-R field strengths is initial one multiplied by imaginary unit \cite{timelike,bnbstip2}. Using double space enables to unify all three theories. This could be a way toward better understanding M-theory.

The basic presumption for implementing Buscher T-dualization procedure is existence of global isometry along some directions. Effectively,
it means that we can find the coordinate basis in which background fields do not depend on those directions \cite{B,RV,GPR,AABL,nasnpb,englezi}.

Except the standard Buscher procedure, there is a generalized Buscher procedure dealing with T-dualization along directions on which background fields depend on. In the generalized Buscher procedure, comparing with the standard one, an additional ingredient is present  and that is invariant coordinate, $x^\mu_{inv}=\int d\xi^\alpha D_\alpha x^\mu$,
where $D_\alpha$ is world-sheet covariant derivative.
So far the generalized procedure was applied in two cases: bosonic string moving in the weakly curved background \cite{DS1,DNS2,DNS} and case where metric is quadratic in coordinates and Kalb-Ramond field
is linear function of coordinates \cite{DNS3}. In the first case isometry is not obvious but actually exists, while in the second case isometry is absent.

Buscher T-dualization procedure was used in the papers \cite{Lust,ALLP,ALLP2,L,nongeo1} in the context
of closed string noncommutativity. In these articles it was considered the coordinate dependent background - constant metric and Kalb-Ramond field with only one nonzero component, $B_{xy}=Hz$, where
the field strength $H$ is infinitesimal.

Buscher T-dualization procedure can be considered as definition of T-dualization. But there is one picturesque way of T-duality representation using double space and permutation group. The name "double" comes from the way how it is constructed. Double space coordinate $Z^M$ consists of initial coordinates $x^\mu$ and their T-dual ones, $y_\mu$, $Z^M=(x^\mu,y_\mu)$ $(\mu=0,1,2,\dots,D-1)$. The formalism emerged about twenty years ago and it was addressed in Refs.\cite{Duff,AAT1,AAT2,WS1,WS2}. In recent years the interest for this formalism is revived \cite{Hull,Hull2,berman,negeom,hohmz}. In these articles T-duality is related with $O(d,d)$ transformations. On the other hand, in Refs.\cite{Duff,sazdam,sazda,bnbstip2}, T-dualization along some subset of directions is represented as permutation of that subset of initial coordinates and the corresponding T-dual ones. T-duality becomes symmetry transformation in double space.

In the article \cite{bnbstip2} we demonstrated the equivalence of the Buscher approach and double space one for type II superstring theory. 
But there is one detail which has to be emphasized. In the mentioned article we used the pure spinor type II superstring action with constant background fields in the form of the first order theory 
i.e. fermionic momenta, $\pi_\alpha$ and $\bar\pi_\alpha$, are not integrated out. In that case the process of T-dualization is mathematically simple, but the price is that the T-dual 
R-R field strength $P^{\alpha\beta}$ could not be obtained within double space formalism. The reason is that R-R field strength is coupled only with the fermionic degrees of freedom which are not dualized. To reproduce Buscher form of the T-dual R-R field strength we made some additional assumptions.

In this article we will integrate out the momenta and obtain theory in terms of the derivatives of the bosonic, $x^\mu$, and fermionic, $\theta^\alpha$ and $\bar\theta^\alpha$, coordinates. After fermionic momenta are integrated out, R-R field strength is coupled with $\partial_\pm x^\mu$. It turns that Buscher T-dualization with such an action is slightly more complicated, but, as it is expected, gives the same result as in the case of the first order theory. The mathematical framework for double space T-dualization is the same as in \cite{bnbstip2}.
We rewrite the T-dual transformation laws
in terms of the double space coordinates $Z^M$ introducing: the generalized metric $\check{\cal H}_{MN}$, the generalized current $\check J_{\pm M}$ and permutation matrix ${\cal T}^M{}_N$, which
swaps the initial coordinates $x^\mu$ and T-dual ones $y_\mu$. Demanding that T-dual double space coordinates, ${}^\star Z^M={\cal T}^M{}_N Z^N$, satisfy the transformation law of the same form as initial coordinates, $Z^M$, we obtain the expressions for T-dual generalized metric, ${}^\star \check{\mathcal H}_{MN}=({\mathcal T} \check{\mathcal H}{\mathcal T})_{MN}$, and T-dual current, ${}^\star \check J_{\pm M}=({\cal T} \check J_{\pm})_M$. 

There is an advantage when we perform T-duality within double space formalism. The main benefit of the using the action with integrated fermionic momenta is that we 
 get all T-dual background fields.


\section{Buscher T-dualization of type II superstring theory with integrated fermionic momenta}
\setcounter{equation}{0}

In this section we will introduce type II superstring action in pure spinor formulation \cite{berko,susyNC,NPBref}
in the approximation of constant background fields and up to the quadratic terms. Then we will integrate out fermionic momenta and 
apply standard Buscher procedure. This leads to more complicated calculations, but in double space an advantage occurs. 

\subsection{Type II superstring in pure spinor formulation}

The general form of the action is borrowed from \cite{verteks} and it is of the form
\begin{equation}\label{eq:vsg}
S=\int_\Sigma d^2 \xi (X^T)^M A_{MN}\bar X^N+S_\lambda+S_{\bar\lambda}\, ,
\end{equation}
where vectors $X^M$ and $\bar X^N$ are left and right chiral supersymetric variables
\begin{equation}
X^M=\left(
\begin{array}{c}
\partial_+ \theta^\alpha\\\Pi_+^\mu\\d_\alpha\\\frac{1}{2}N_+^{\mu\nu}
\end{array}\right)\, ,\quad \bar X^M=\left(
\begin{array}{c}
\partial_-\bar\theta^\alpha\\\Pi_-^\mu\\\bar d_\alpha\\\frac{1}{2}\bar N_-^{\mu\nu}
\end{array}\right),
\end{equation}
which components are defined as
\begin{equation}
\Pi_+^\mu=\partial_+ x^\mu+\frac{1}{2}\theta^\alpha (\Gamma^\mu)_{\alpha\beta}\partial_+ \theta^\beta\, ,\quad \Pi_-^\mu=\partial_- x^\mu+\frac{1}{2}\bar\theta^\alpha (\Gamma^\mu)_{\alpha\beta}\partial_- \bar\theta^\beta\, ,
\end{equation}
\begin{eqnarray}
d_\alpha&=&\pi_\alpha-\frac{1}{2}(\Gamma_\mu \theta)_\alpha\left[ \partial_+ x^\mu +\frac{1}{4} (\theta \Gamma_\mu \partial_+\theta)\right]\, ,\nonumber\\
\bar d_\alpha&=&\bar\pi_\alpha-\frac{1}{2}(\Gamma_\mu \bar\theta)_\alpha \left[\partial_- x^\mu +\frac{1}{4} (\bar\theta \Gamma_\mu \partial_-\bar\theta)\right]\, ,
\end{eqnarray}
\begin{equation}\label{eq:Npm}
N_+^{\mu\nu}=\frac{1}{2}w_\alpha(\Gamma^{[\mu\nu]})^\alpha{}_\beta \lambda^\beta\, ,\quad \bar N_-^{\mu\nu}=\frac{1}{2}\bar w_\alpha (\Gamma^{[\mu\nu]})^\alpha{}_\beta \bar\lambda^\beta\, .
\end{equation}
The supermatrix $A_{MN}$ is of the form
\begin{equation}\label{eq:Amn}
A_{MN}=\left(\begin{array}{cccc}
A_{\alpha\beta} & A_{\alpha\nu} & E_\alpha{}^\beta & \Omega_{\alpha,\mu\nu}\\
A_{\mu\beta} & A_{\mu\nu} & \bar E_\mu^\beta & \Omega_{\mu,\nu\rho}\\
E^\alpha{}_\beta & E^\alpha_\nu & {\rm P}^{\alpha\beta} & C^\alpha{}_{\mu\nu}\\
\Omega_{\mu\nu,\beta} & \Omega_{\mu\nu,\rho} & \bar C_{\mu\nu}{}^\beta & S_{\mu\nu,\rho\sigma}
\end{array}\right)\, .
\end{equation}

The world sheet $\Sigma$ is parameterized by
$\xi^m=(\xi^0=\tau\, ,\xi^1=\sigma)$ and
$\partial_\pm=\partial_\tau\pm\partial_\sigma$. Superspace is spanned by bosonic coordinates $x^\mu$ ($\mu=0,1,2,\dots,9$) and fermionic ones $\theta^\alpha$ and $\bar\theta^{\alpha}$
$(\alpha=1,2,\dots,16)$. The variables $\pi_\alpha$ and $\bar
\pi_{\alpha}$ are canonically conjugated momenta to
$\theta^\alpha$ and $\bar\theta^\alpha$, respectively. The actions for pure spinors, $S_\lambda$ and $S_{\bar\lambda}$, are free field actions
\begin{equation}
S_\lambda=\int d^2\xi w_\alpha \partial_-\lambda^\alpha\, ,\quad S_{\bar\lambda}=\int d^2\xi \bar w_\alpha \partial_+ \bar\lambda^\alpha\, ,
\end{equation}
where $\lambda^\alpha$ and $\bar\lambda^\alpha$ are pure spinors and $w_\alpha$ and $\bar w_\alpha$ are their canonically conjugated momenta, respectively. The pure spinors satisfy so called pure spinor constraints
\begin{equation}\label{eq:psc0}
\lambda^\alpha (\Gamma^\mu)_{\alpha\beta}\lambda^\beta=\bar\lambda^\alpha (\Gamma^\mu)_{\alpha\beta}\bar\lambda^\beta=0\, .
\end{equation}
This action (\ref{eq:vsg}) for type II superstring in pure spinor formulation is general one and it is constructed as an expansion in powers of $\theta^\alpha$ and $\bar\theta^\alpha$ (for details see \cite{verteks}).

Our plan is to implement full T-dualization which means that we T-dualize along all bosonic directions $x^\mu$. Consequently, we will assume that background fields do not depend on them. On the other hand, because
of the way how the action is constructed, for practical reasons (mathematical simplification), we will consider just the first components in the expansion in powers of $\theta^\alpha$ and $\bar\theta^\alpha$. Effectively, this means that nonzero background fields
are constant. The background fields from the first and last columns and rows in the matrix $A_{MN}$ are zero (detailed explanation could be found in \cite{verteks,bnbstip2}). The fields surviving these approximations are known in literature as physical superfields because their first components are supergravity fields.

Finally, all our assumptions produce
\begin{equation}
\Pi_\pm^\mu\to \partial_{\pm} x^\mu\, ,\quad d_\alpha\to \pi_\alpha\, ,\quad \bar d_\alpha\to \bar\pi_\alpha\, ,
\end{equation}
where physical superfields take form
\begin{equation}
A_{\mu\nu}=\kappa(\frac{1}{2}g_{\mu\nu}+B_{\mu\nu})+\frac{1}{4\pi}\eta_{\mu\nu}\Phi\, ,\quad E^\alpha_\nu=-\Psi^\alpha_\nu\, ,\quad \bar E_\mu^\alpha=\bar\Psi_\mu^\alpha\, ,\quad {\rm P}^{\alpha\beta}=\frac{1}{2\kappa}P^{\alpha\beta}\, .
\end{equation}
Here $g_{\mu\nu}$ is symmetric and $B_{\mu\nu}$ is antisymmetric
tensor.
Consequently, the full action $S$ is
\begin{eqnarray}\label{eq:SB}
&{}&S=\kappa \int_\Sigma d^2\xi \left[\partial_{+}x^\mu
\Pi_{+\mu\nu}\partial_- x^\nu+\frac{1}{4\pi\kappa}\Phi R^{(2)}\right] \\&+&\int_\Sigma d^2 \xi \left[
-\pi_\alpha
\partial_-(\theta^\alpha+\Psi^\alpha_\mu
x^\mu)+\partial_+(\bar\theta^{\alpha}+\bar \Psi^{\alpha}_\mu
x^\mu)\bar\pi_{\alpha}+\frac{1}{2\kappa}\pi_\alpha P^{\alpha
\beta}\bar \pi_{\beta}\right ]\, ,\nonumber
\end{eqnarray}
where $G_{\mu\nu}=\eta_{\mu\nu}+g_{\mu\nu}$ is metric tensor and
\begin{equation}\label{eq:pipm1}
\Pi_{\pm \mu\nu}=B_{\mu\nu}\pm \frac{1}{2}G_{\mu\nu}\, .
\end{equation}
We will neglect the Tseytlin term in the further analysis because, for constant dilaton field $\Phi$, it is proportional to the Euler caracteristic. Consequently, on some given manifold that term is constant.
Actions $S_\lambda$ and $S_{\bar\lambda}$ are decoupled from the rest and the action, in its final form, is  ghost independent.

\subsection{Full bosonic T-dualization using Buscher rules}

Let us now integrate out fermionic momenta from action (\ref{eq:SB}) and obtain the theory expressed in terms of the supercoordinates $(x^\mu,\theta^\alpha,\bar\theta^\alpha)$ and 
their world-sheet derivatives. On the equations of motion for fermionic momenta $\pi_\alpha$ and $\bar\pi_\alpha$,
\begin{equation}\label{eq:impulsi}
\pi_\alpha=-2\kappa\partial_+\left(\bar\theta^\beta+\bar\Psi^\beta_\mu x^\mu\right)(P^{-1})_{\beta\alpha}\, ,\quad \bar\pi_\alpha=2\kappa(P^{-1})_{\alpha\beta}\partial_-\left(\theta^\beta+\Psi^\beta_\mu x^\mu\right)\, ,
\end{equation}
the action gets the form
\begin{eqnarray}\label{eq:lcdejstvo}
&{}&S=\kappa \int_\Sigma d^2 \xi \partial_+ x^\mu \left[\Pi_{+\mu\nu}+2\bar\Psi^\alpha_\mu (P^{-1})_{\alpha\beta}\Psi^\beta_\nu\right]\partial_- x^\nu+\\
&+&2\kappa \int_\Sigma d^2 \xi \left[\partial_+ \bar\theta^\alpha (P^{-1})_{\alpha\beta}\partial_-\theta^\beta +\partial_+ \bar\theta^\alpha (P^{-1}\Psi)_{\alpha\mu}\partial_-x^\nu+\partial_+x^\mu (\bar\Psi P^{-1})_{\mu\alpha}\partial_- \theta^\alpha\right]\nonumber\, .
\end{eqnarray} 

We will perform bosonic T-dualization of the action (\ref{eq:lcdejstvo}) along all directions $x^\mu$ using Buscher T-dualization rules. In order to gauge global symmetry $\delta x^\mu=\lambda^\mu$, we introduce covariant derivatives, $D_\pm x^\mu=\partial_\pm x^\mu+v^\mu_\pm$, instead the ordinary ones, $\partial_\pm x^\mu$, where $v_\pm^\mu$ are gauge fields. Fixing the gauge ($x^\mu=const.$) means effectively that ordinary derivatives  $\partial_\pm x^\mu$ are replaced with gauge fields $v_\pm^\mu$ in the initial action (\ref{eq:lcdejstvo}),
while, in order to make $v_\pm^\mu$ to be unphysical degrees of freedom, we add to the action
\begin{equation}
S_{add}=\frac{\kappa}{2}\int d^2\xi (v_+^\mu \partial_- y_\mu-v_-^\mu \partial_+y_\mu)\, .
\end{equation}
The gauged fixed action is of the form
{\small{\begin{eqnarray}\label{eq:sgf}
&{}&S_{fix}=S+S_{add}\\&=&\kappa\int d^2\xi \left[v_+^\mu \Pi_{+\mu\nu}v_-^\nu+2(\partial_+\bar\theta^\alpha+\bar\Psi^\alpha_\mu v^\mu_+)(P^{-1})_{\alpha\beta}(\partial_-\theta^\beta+\Psi^\beta_\nu v^\nu_-)+\frac{1}{2}(v_+^\mu \partial_- y_\mu-v_-^\mu \partial_+y_\mu)\right]\, .\nonumber
\end{eqnarray}}}
Varying the gauge fixed action (\ref{eq:sgf}) with respect to the Lagrange multipliers $y_\mu$, we obtain that field strength for gauge fields $v_\pm^\mu$ is equal to zero
\begin{equation}\label{eq:vary}
\partial_+ v^\mu_--\partial_- v^\mu_+=0 \Rightarrow v_\pm^\mu=\partial_\pm x^\mu\, .
\end{equation}
In this way we restore initial theory from gauge fixed action.
Let us note that we omitted the dilaton term because this is a classical analysis while dilaton is treated within quantum formalism. 

Varying the gauge fixed action with respect the gauge fields $v^\mu_+$ and $v^\mu_-$, we get the equations, respectively
\begin{equation}
\Pi_{+\mu\nu}v^\nu_-+2\bar\Psi^\alpha{}_\mu (P^{-1})_{\alpha\beta}(\partial_-\theta^\beta+\Psi^\beta{}_\nu v^\nu_-)+\frac{1}{2}\partial_- y_\mu=0\, ,
\end{equation}
\begin{equation}
v^\nu_+ \Pi_{+\nu\mu}+2(\partial_+\bar\theta^\alpha+\bar\Psi^\alpha{}_\nu v^\nu_+)(P^{-1})_{\alpha\beta}\Psi^\beta{}_\mu-\frac{1}{2}\partial_+ y_\mu=0\, .
\end{equation}
Here we introduce the new notation
\begin{equation}
\check \Pi_{+\mu\nu}\equiv \Pi_{+\mu\nu}+2\bar\Psi^\alpha{}_\mu (P^{-1})_{\alpha\beta}\Psi^\beta{}_\nu=\check B_{\mu\nu}+\frac{1}{2}\check G_{\mu\nu}\, ,
\end{equation}
where $\check B_{\mu\nu}$ and $\check G_{\mu\nu}$ are antisymmetric and symmetric parts of $\check\Pi_{+\mu\nu}$, respectively. These expressions are in fact Kalb-Ramond field $B_{\mu\nu}$ and metric $G_{\mu\nu}$ improved by some expressions consisting of the NS-R and R-R background fields.
The above expressions for gauge fields can be rewritten in the form
\begin{equation}
\check \Pi_{+\mu\nu}v^\nu_-+2\bar\Psi^\alpha{}_\mu (P^{-1})_{\alpha\beta}\partial_- \theta^\beta+\frac{1}{2}\partial_- y_\mu=0\, ,
\end{equation}
\begin{equation}
v^\nu_+ \check\Pi_{+\nu\mu}+2\partial_+\bar\theta^\alpha (P^{-1})_{\alpha\beta}\Psi^\beta{}_\mu-\frac{1}{2}\partial_+ y_\mu=0\, .
\end{equation}
Using the relations
\begin{equation}\label{eq:tetapi1}
\check\Theta^{\mu\nu}_- \check\Pi_{+\nu\rho}=\frac{1}{2\kappa}\delta^\mu{}_\rho\, ,\quad \check\Theta^{\mu\nu}_-=-\frac{2}{\kappa}(\check G_E^{-1}\check\Pi_{-}\check G^{-1})^{\mu\nu}\, ,
\end{equation}
where
\begin{equation}
\check\Theta^{\mu\nu}_-=\Theta^{\mu\nu}_-- 4\kappa \Theta^{\mu\rho}_- \bar\Psi^\alpha{}_\rho (\tilde P^{-1})_{\alpha\beta} \Psi^\beta{}_\lambda \Theta^{\lambda\nu}_-=\check\Theta^{\mu\nu}+\frac{1}{\kappa}(\check G_E^{-1})^{\mu\nu}\, ,
\end{equation}
\begin{equation}
\tilde P^{\alpha\beta}\equiv P^{\alpha\beta}+4\kappa\Psi^\alpha{}_\mu \Theta^{\mu\nu}_- \bar\Psi^\beta{}_\nu\, ,
\end{equation}
\begin{equation}
\Theta_-^{\mu\nu}=-\frac{2}{\kappa}(G_E^{-1}\Pi_- G^{-1})^{\mu\nu}\, ,\quad \Theta^{\mu\rho}_- \Pi_{+\rho\nu}=\frac{1}{2\kappa}\delta^\mu{}_\nu\, .
\end{equation}
we get
\begin{equation}\label{eq:v-k}
v^\mu_-=-\kappa \check\Theta^{\mu\nu}_- \partial_-\left[y_\nu+4\bar\Psi^\alpha{}_\mu (P^{-1})_{\alpha\beta} \theta^\beta\right]\, ,
\end{equation}
\begin{equation}\label{eq:v+k}
v^\mu_+=\kappa \partial_+\left[y_\nu-4\bar\theta^\alpha(P^{-1})_{\alpha\beta}\Psi^\beta{}_\nu\right]\check\Theta^{\nu\mu}_-\, .
\end{equation}
The relation (\ref{eq:tetapi1}) is proved by direct calculation and using the definition of $\tilde P^{\alpha\beta}$.

Inserting the expressions (\ref{eq:v-k}) and (\ref{eq:v+k}) into the expression for gauge fixed action (\ref{eq:sgf}) we obtain T-dual action
\begin{eqnarray}
&{}&{}^\star S=\kappa \int d^2\xi \left[ \frac{\kappa}{2}\partial_+ y_\mu \check\Theta^{\mu\nu}_- \partial_-y_\nu+2\kappa \partial_+y_\mu \check\Theta^{\mu\nu}_- \bar\Psi^\alpha{}_\nu (P^{-1})_{\alpha\beta}\partial_-\theta^\beta\right.\\&-&\left.2\kappa \partial_+ \bar\theta^\alpha (P^{-1})_{\alpha\beta}\Psi^\beta{}_\mu \check\Theta^{\mu\nu}_- \partial_- y_\nu+2\partial_+\bar\theta^\alpha(P^{-1}-4\kappa P^{-1}\Psi \check\Theta_- \bar\Psi P^{-1})_{\alpha\beta}\partial_-\theta^\beta\right]\, .\nonumber
\end{eqnarray}
Introducing T-dual background fields marked by ${}^\star$, we write the T-dual action in the form of the initial action (\ref{eq:lcdejstvo})
\begin{eqnarray}
&{}&{}^\star S=\kappa \int d^2\xi \left[ \partial_+ y_\mu\left({}^\star\Pi_++2{}^\star\bar\Psi {}^\star P^{-1}{}^\star \Psi\right)^{\mu\nu}\partial_- y_\nu+2\partial_+ y_\mu ({}^\star\bar\Psi {}^\star P^{-1})^\mu{}_\alpha \partial_-\theta^\alpha\right.\nonumber\\
&+& \left. 2\partial_+ \bar\theta^\alpha ({}^\star P^{-1}{}^\star\Psi)_\alpha{}^\mu \partial_-y_\mu+2\partial_+\bar\theta^\alpha ({}^\star P^{-1})_{\alpha\beta}\partial_-\theta\right]\, .
\end{eqnarray}
Comparing last two equations, we get the T-dual background fields in terms of the initial ones
\begin{equation}\label{eq:bozpo1}
{}^\star\Pi_+^{\mu\nu}+2{}^\star\bar\Psi^{\mu\alpha} ({}^\star P^{-1})_{\alpha\beta} {}^\star \Psi^{\beta\nu}=\frac{\kappa}{2}\check\Theta^{\mu\nu}_-\, ,
\end{equation}
\begin{equation}\label{eq:tdualpsi}
{}^\star\bar\Psi^{\mu\beta} ({}^\star P^{-1})_{\beta\alpha}=\kappa \check\Theta^{\mu\nu}_- \bar\Psi^\beta{}_\nu (P^{-1})_{\beta\alpha}\, ,\quad ({}^\star P^{-1})_{\alpha\beta}{}^\star \Psi^{\beta\mu}=-\kappa (P^{-1})_{\alpha\beta}\Psi^\beta{}_\nu \check \Theta^{\nu\mu}_-\, ,
\end{equation}
\begin{equation}\label{eq:dualp}
({}^\star P^{-1})_{\alpha\beta}=(P^{-1})_{\alpha\beta}-4\kappa (P^{-1})_{\alpha\gamma}\Psi^\gamma{}_\mu \check\Theta^{\mu\nu}_- \bar\Psi^\delta{}_\nu (P^{-1})_{\delta\beta}\, .
\end{equation}
By direct calculation, solving above four equations, we get finally
\begin{equation}\label{eq:fintdpi}
{}^\star \Pi^{\mu\nu}_{+}=\frac{\kappa}{2}\Theta^{\mu\nu}_-\, ,
\end{equation}
\begin{equation}
{}^\star \Psi^{\alpha\mu}=-\kappa \Psi^\alpha{}_\nu \Theta_-^{\nu\mu}\, ,\quad {}^\star \bar\Psi^{\mu\alpha}=\kappa \Theta^{\mu\nu}_- \bar\Psi^\alpha{}_\nu\, ,
\end{equation}
\begin{equation}
{}^\star P^{\alpha\beta}=\tilde P^{\alpha\beta}\, ,
\end{equation}
which is in full agreement with the case where we T-dualize the same model in the form of the first order theory. 

Combining equations of motion for Lagrange multiplier (\ref{eq:vary}) with the equations of motion for gauge fields, (\ref{eq:v-k}) and (\ref{eq:v+k}), we obtain the relation between initial $x^\mu$ and T-dual coordinates $y_\mu$
\begin{equation}\label{eq:tdlawx}
\partial_\pm x^\mu \cong -\kappa \check \Theta_\pm^{\mu\nu} \left[\partial_\pm y_\nu+4\Psi_\pm^\alpha{}_\nu (P_\mp^{-1})_{\alpha\beta}\partial_\pm \theta^\beta_\mp\right]\, .
\end{equation}
Inverse of this relation is also useful and it is of the form
\begin{equation}\label{eq:tdlawy}
\partial_\pm y_\mu\cong -2\check \Pi_{\mp\mu\nu}\partial_\pm x^\nu-4\Psi^\alpha_{\pm \mu} (P_{\mp}^{-1})_{\alpha\beta} \partial_\pm \theta^\beta_\mp\, .
\end{equation}
Here we use the notation
\begin{equation}\label{eq:not1}
\theta_+^\alpha\equiv \theta^\alpha\, , \quad\theta_-^\alpha\equiv \bar\theta^\alpha\, ,
\end{equation}
\begin{equation}\label{eq:not2}
P_+^{\alpha\beta}\equiv P^{\alpha\beta}\, ,\quad P_-^{\alpha\beta}\equiv P^{\beta\alpha}\, ,
\end{equation}
\begin{equation}\label{eq:not3}
\Psi_{+\mu}^\alpha\equiv \Psi^\alpha{}_\mu\, ,\quad \Psi^\alpha_{-\mu}\equiv \bar\Psi^\alpha{}_\mu\, ,
\end{equation}
\begin{equation}\label{eq:not4}
\check\Theta_+^{\mu\nu}\equiv -\check\Theta^{\nu\mu}_-\, .
\end{equation}

As we see two chirality sectors transform differently under T-dualization. The form of T-dualization transformation laws is of the same form as in \cite{bnbstip2}. Consequently, in accordance with the results of the papers \cite{H,BPT,bnbstip2}, we introduce proper fermionic coordinates
\begin{equation}
{}^\bullet \theta^\alpha_+=\theta^\alpha_+\, ,\quad {}^\bullet \bar\theta^\alpha_-=-(\Gamma_{11} \bar\theta_-)^\alpha\, , 
\end{equation}
and the correct form of T-dual fields is
\begin{equation}\label{eq:not5}
{}^\star \Pi^{\mu\nu}_{+}=\frac{\kappa}{2}\Theta^{\mu\nu}_-\, ,
\end{equation}
\begin{equation}
{}^\star \Psi^{\alpha\mu}=-\kappa \Psi^\alpha{}_\nu \Theta_-^{\nu\mu}\, ,\quad {}^\star \bar\Psi^{\mu\alpha}=-\kappa \Theta^{\mu\nu}_- (\Gamma_{11}\bar\Psi )^\alpha{}_\nu\, ,
\end{equation}
\begin{equation}\label{eq:not7}
{}^\star P^{\alpha\beta}=-(\tilde P \Gamma_{11})^{\alpha\beta}\, .
\end{equation}

\section{T-dualization of type II superstring in double space}
\setcounter{equation}{0}

In this section we will demonstrate another framework in which we can perform T-dualization procedure. Unlike the case of the T-dualization of type II superstring theory in the form of the first order theory \cite{bnbstip2} where T-dual R-R field strength is not obtained within double space framework, here we will see that, when fermionic momenta are integrated out, double space formalism gives all T-dual background fields. Before the T-dualization procedure we will introduce double space and corresponding quantities.

\subsection{T-dual transformation law in double space}

Let us introduce the double space coordinate
\begin{equation}
Z^M=\left(
\begin{array}{c}
x^\mu\\y_\mu
\end{array}\right)\, ,
\end{equation}
and rewrite the T-dual transformation laws (\ref{eq:tdlawx})-(\ref{eq:tdlawy}) in the form
\begin{equation}\label{eq:tdualc1}
\pm\partial_\pm y_\mu\cong \check G^E_{\mu\nu}\partial_\pm x^\nu+\kappa \check G^E_{\mu\rho}\check \Theta^{\rho \nu} \partial_\pm y_\nu +4\kappa G^E_{\mu\rho}\Theta_{\pm}^{\rho\lambda}\Psi_{\pm \lambda}^\alpha (P_{\mp}^{-1})_{\alpha\beta}\partial_\pm \theta_\mp^\beta\, ,
\end{equation}
\begin{equation}\label{eq:tdualc2}
\pm \partial_{\pm} x^\mu\cong (\check G^{-1})^{\mu\nu}\partial_\pm y_\nu+2(\check G^{-1}\check B)^\mu{}_\nu \partial_\pm x^\nu+4(\check G^{-1})^{\mu\rho}\Psi_{\pm \rho}^\alpha (P_{\mp}^{-1})_{\alpha\beta}\partial_\pm \theta_\mp^\beta\, .
\end{equation}
These two equations can be rewritten in double space as
\begin{equation}\label{eq:tdlds}
\pm \Omega_{MN}\partial_\pm Z^N \cong \check {\mathcal H}_{MN}\partial_\pm Z^N+\check J_{\pm M}\, ,
\end{equation}
where the generalized metric is of the form
\begin{equation}
\check {\mathcal H}_{MN}=\left(
\begin{array}{cc}
\check G^E_{\mu\nu} & \kappa \check G^E_{\mu\rho}\check \Theta^{\rho\nu}\\
2(\check G^{-1})^{\mu\rho}\check B_{\rho\nu} & (\check G^{-1})^{\mu\nu}
\end{array}\right)\, ,
\end{equation}
and the double current is
\begin{equation}
\check J_{\pm M}=4\left(
\begin{array}{c}
\kappa \check G^E_{\mu\rho}\check \Theta^{\rho\nu}_\pm\\
(\check G^{-1})^{\mu\nu}
\end{array}\right) J_{\pm \mu}\, ,\quad J_{\pm \mu}=\Psi^\alpha_{\pm \mu}(P_\mp^{-1})_{\alpha\beta} \partial_\pm \theta_\mp^\beta\, .
\end{equation}
The matrix
\begin{equation}
\Omega=\left(
\begin{array}{cc}
0 & 1_D\\
1_D & 0
\end{array}\right)\, ,
\end{equation}
where $1_D$ denotes the unity matrix in $D$ dimensions, is known in double field theory (DFT) as invariant $SO(D,D)$ metric. 

Note that generalized metric is not of the standard form because its components contain improved Kalb-Ramond field $\check B_{\mu\nu}$ and improved metric $\check G_{\mu\nu}$. Those additional factors in $\check B_{\mu\nu}$ and $\check G_{\mu\nu}$ have bilinear form in NS-R fields $\Psi^\alpha{}_\mu$ and $\bar\Psi^\alpha{}_\mu$.
Still, it holds
\begin{equation}
\check {\mathcal H}^T \Omega \check {\mathcal H}=\Omega\, , \quad \Omega^2=1\, ,\quad \det {\check{\mathcal H}}=1\, ,
\end{equation}
which means that $\check{\mathcal H}\in SO(D,D)$.

\subsection{Full T-dualization in double space}

Let us introduce the permutation matrix
\begin{equation}
\mathcal T^M{}_N=\left(
\begin{array}{cc}
0 & 1_D\\
1_D & 0
\end{array}\right)\, ,
\end{equation}
and define T-dual double coordinate ${}^\star Z^M$ as
\begin{equation}
 {}^\star Z^M=\mathcal T^M{}_N Z^N\, .
\end{equation}

We demand that T-dual transformation law for T-dual double coordinate ${}^\star Z^M$ has the same form as for initial double coordinate $Z^M$ (\ref{eq:tdlds})
\begin{equation}
\pm \Omega_{MN}\partial_\pm {}^\star Z^N\cong {}^\star{\check\mathcal H}_{MQ}\partial_\pm {}^\star Z^Q+{}^\star \check J_{\pm M}\, ,
\end{equation}
which implies that T-dual generalized metric and double current are of the form, respectively,
\begin{equation}\label{eq:tdjed1}
{}^\star{\check\mathcal H}_{MN}={\mathcal T}_M{}^P {\check\mathcal H}_{PQ} {\mathcal T}^Q{}_N\, ,\quad {}^\star {\check J}_{\pm M}=\mathcal T_M{}^N \check J_{\pm N\, .}
\end{equation}

Let us make explicit the first equation in (\ref{eq:tdjed1})
\begin{equation}
\left(
\begin{array}{cc}
{}^\star \check G_E^{\mu\nu} & \kappa {}^\star \check G_E^{\mu\rho}{}^\star \check\Theta_{\rho\nu}\\
2 ({}^\star\check G^{-1})_{\mu\rho}{}^\star \check B^{\rho\nu} & ({}^\star\check G^{-1})_{\mu\nu}
\end{array}\right)=\left(
\begin{array}{cc}
(\check G^{-1})^{\mu\nu} & 2(\check G^{-1})^{\mu\rho} \check B_{\rho\nu}\\
\kappa \check G^E_{\mu\rho}\check \Theta^{\rho\nu} & \check G^E_{\mu\nu}
\end{array}\right)\, .
\end{equation}
Equating $(2,2)$ block-components we get
\begin{equation}\label{eq:com22}
({}^\star \check G^{-1})_{\mu\nu}=\check G^E_{\mu\nu}\, ,
\end{equation}
which produces
\begin{equation}
{}^\star \check G^{\mu\nu}=(\check G_E^{-1})^{\mu\nu}\, .
\end{equation}
Using $(2,1)$ block components equation
\begin{equation}
2 ({}^\star\check G^{-1})_{\mu\rho}{}^\star \check B^{\rho\nu}=\kappa \check G^E_{\mu\rho}\check \Theta^{\rho\nu}\, ,
\end{equation}
and combining with (\ref{eq:com22}). we obtain
\begin{equation}
{}^\star \check B^{\mu\nu}=\frac{\kappa}{2}\check \Theta^{\mu\nu}\, .
\end{equation}
Using these two results we have
\begin{equation}\label{eq:dualpi}
{}^\star \check\Pi^{\mu\nu}_{\pm}={}^\star B^{\mu\nu}\pm \frac{1}{2}{}^\star \check G^{\mu\nu}=\frac{\kappa}{2}\left[\check\Theta^{\mu\nu}\pm \frac{1}{\kappa}(\check G_E^{-1})^{\mu\nu}\right]=\frac{\kappa}{2}\check\Theta^{\mu\nu}_{\mp}\, .
\end{equation}
Obtained equation coincides with the equation obtained by standard Buscher procedure (\ref{eq:bozpo1}). The block-components $(1,1)$ and $(1,2)$ give, respectively,
\begin{equation}
{}^\star \check G_E^{\mu\nu}=(\check G^{-1})^{\mu\nu}\, ,\quad {}^\star \check G_E^{\mu\rho}{}^\star \check\Theta_{\rho\nu}=\frac{2}{\kappa}(\check G^{-1})^{\mu\rho}\check B_{\rho\nu}\, ,
\end{equation}
Combining last two equations produces
\begin{equation}
{}^\star \check \Theta_{\mu\nu}=\frac{2}{\kappa} \check B_{\mu\nu}\, ,
\end{equation}
and, further, we have
\begin{equation}\label{eq:dualteta}
{}^\star \check\Theta_{-\mu\nu}=\frac{2}{\kappa}\check\Pi_{+\mu\nu}\, .
\end{equation}

Using the relations between initial and T-dual NS-NS background fields obtained above and the second equation in (\ref{eq:tdjed1}), we get
\begin{equation}\label{eq:jmustruje}
{}^\star J^{\mu}_\pm=\kappa \check \Theta^{\mu\nu}_\pm J_{\pm \nu}\, ,
\end{equation}
where T-dual current ${}^\star J^\mu_\pm$ has the same form as initial one but in terms of T-dual background fields and proper fermionic coordinates (for details see \cite{bnbstip2})
\begin{equation}\label{eq:properstruja}
{}^\star J^\mu_\pm\equiv {}^\star \Psi_{\pm}^{\alpha\mu}({}^\star P_\mp^{-1})_{\alpha\beta} \partial_\pm {}^\bullet\theta^\beta_\mp\, .
\end{equation}
Proper fermionic coordinates are defined as
\begin{equation}
{}^\bullet\theta_+^\alpha=\theta_+^\alpha\, ,\quad {}^\bullet\theta_-^\alpha=-(\Gamma_{11}\theta_-)^\alpha\, ,
\end{equation}

From the relation (\ref{eq:properstruja}) we have
\begin{equation}\label{eq:starcekpi}
{}^\star \check\Pi_+^{\mu\nu}={}^\star\Pi_+^{\mu\nu}+2{}^\star \bar\Psi^{\alpha\mu} ({}^\star P^{-1})_{\alpha\beta}{}^\star \Psi^{\beta\nu}\, , 
\end{equation}
\begin{equation}\label{eq:starcekteta}
{}^\star \check\Theta_{-\mu\nu}={}^\star \Theta_{-\mu\nu}-4\kappa {}^\star \Theta_{-\mu\rho} {}^\star \bar\Psi^{\rho\alpha} ({}^\star \tilde P^{-1})_{\alpha\beta}{}^\star \Psi^{\beta\lambda}{}^\star \Theta_{-\lambda\nu}\, ,
\end{equation}
and, solving these equations, we get
\begin{eqnarray}\label{eq:poljakapa}
&{}&{}^\star \Psi^{\alpha\mu}=\pm\kappa \Psi^{\alpha}{}_\nu \Theta^{\nu\mu}_-\, ,\quad {}^\star \bar\Psi^{\alpha\mu}=\pm \kappa \Theta_-^{\mu\nu}(\Gamma_{11} \bar\Psi)^\alpha_\nu\, ,\nonumber \\&{}& {}^\star P^{\alpha\beta}=-(P^{\alpha\gamma}+4\kappa \Psi^\alpha{}_\mu \Theta^{\mu\nu}\bar\Psi^\gamma{}_\nu)(\Gamma_{11})_\gamma{}^\beta\, . 
\end{eqnarray}
Here double space formalism produces (\ref{eq:dualpi}) and (\ref{eq:poljakapa}) i.e. all relations (\ref{eq:not5})-(\ref{eq:not7}), up to the sign of the T-dual NS-R background fields. This uncertainty in sign is consequence of the fact that in both equations, (\ref{eq:starcekpi}) and (\ref{eq:starcekteta}), NS-R fields, ${}^\star \Psi^{\alpha\mu}$ and ${}^\star \bar\Psi^{\alpha\mu}$, appear in bilinear combination. In comparison with the case where fermionic momenta are not integrated out \cite{bnbstip2}, this is improvement, because double formalism gives all T-dual background fields. In Ref.\cite{bnbstip2} we did not obtain the relation for T-dual R-R background field and we had to impose additional conditions. Here calculation is slightly more complicated, but we obtain all fields within one formalism. The reason is that in \cite{bnbstip2} R-R field is coupled by fermionic momenta which are not T-dualized. Consequently, after integration of fermionic momenta, there appears coupling between R-R field strength with bosonic coordinates $x^\mu$ which results in Eqs.(\ref{eq:poljakapa}). 


\section{Conclusion}
\setcounter{equation}{0}

In this article we considered the type II superstring theory in pure spinor formulation with constant background fields. We integrated out the fermionic momenta and obtained the theory quadratic in world-sheet derivatives of bosonic and fermionic coordinates. Our goal was to show advantage of the T-dualization within double space formalism comparing to the first order theory \cite{bnbstip2}.

At the beginning we explained how we obtained the action with constant background field from the general one derived in \cite{verteks}. The assumed shift symmetry along bosonic directions $x^\mu$ means that background fields do not depend on $x^\mu$. On the other hand, for technical simplicity of the calculations, we take just first terms in the expansions of background fields in powers of $\theta^\alpha$ and $\bar\theta^\alpha$. All these assumptions result in the constant background fields. In the final form of the action just physical superfields are present, while auxiliary fields and field strengths are zero.

The main mathematical difference comparing with the Ref.\cite{bnbstip2} is that fermionic momenta are integrated out. In this way we obtained theory which is quadratic in the world-sheet derivatives of the coordinates, $\partial_\pm x^\mu$, $\partial_\pm \theta^\alpha$ and $\partial_\pm \bar\theta^\alpha$. It is important to emphasize that in such formulation R-R field strength $P^{\alpha\beta}$ is coupled with the derivatives of bosonic coordinates $\partial_\pm x^\mu$. Then we applied Buscher procedure and, beside some slightly more complicated mathematical calculations, we obtained the same result as in the case for the first order theory \cite{bnbstip2}.

Our contribution was to show benefit in performing T-dualization procedure in double space using action (\ref{eq:lcdejstvo}) comparing with the results obtained for action (\ref{eq:SB}) in Ref.\cite{bnbstip2}.

The double space is spanned by coordinates $Z^M=(x^\mu,y_\mu)$, where $x^\mu$ are initial bosonic coordinates and $y_\mu$ are corresponding 
the T-dual ones. The T-dual transformation laws are rewritten in terms of the double space coordinates introducing the generalized metric ${\check {\cal H}}_{MN}$ and the current $\check J_{\pm M}$. Note that their components are expressed in terms of the improved Kalb-Ramond field and metric containing additional terms bilinear in NS-R background fields $\Psi^\alpha_\mu$ and $\bar\Psi^\alpha_\mu$.
Demanding that T-dual double space coordinates ${}^\star Z^M={\cal T}^M{}_N Z^N$
satisfy the transformation law of the same form as the initial coordinates $Z^M$ we found the
T-dual generalized metric ${}^\star {\check {\cal H}}_{MN}$ and the T-dual current ${}^\star J_{\pm M}$. T-dual generalized metric should have the same form as the initial ones, so, in this way we obtain relations which produces the expressions for T-dual background fields in terms of the initial ones, which agrees with that obtained applying Buscher procedure.

In the article \cite{bnbstip2} we obtained the expressions for T-dual NS-NS background fields as well as for NS-R fields. But because we T-dualized along bosonic directions which are not coupled with R-R field strength, double space formalism did not give us the expression for T-dual R-R field strength. These expressions was obtained under some additional assumptions out of double space formalism.

Here we succeeded to obtain expressions for all T-dual background fields which showed  that there is advantage to perform double space T-dualization in the second order theory  where fermionic momenta are integrated out. 

After summarizing the results of this article it is interesting to discuss their significance and relation to the results of other articles addressing the same or similar subjects. 
For example, in the article \cite{rad1} authors construct the type II model with T-duality as manifest symmetry. The way of construction is partially similar to the one 
from \cite{verteks} used in this paper. In \cite{verteks} they used (anti)holomorphicity and nilpotency conditions, while in \cite{rad1} instead nilpotency conditions they used conditions 
originating from $\kappa$ symmetry. But in \cite{rad1} they do not study the problem of RR field strength as well as interchange between typeIIA/B in T-dualization process, which are the subjects 
addressed in this article. Further, in \cite{rad2} one version of the doubled superspace is discussed. It is pretty similar to the space we used in this article, but it is obtained by multiplication 
of the left and right chiral sectors with $\mathcal N=1$ supersymmetry in $D=10$. The space which is obtained is spanned by initial bosonic coordinates, their T-dual ones and two fermionic coordinates. 
Our doubled coordinate contains just bosonic part of this doubled supercoordinate because we consider here just bosonic T-dualization and, consequently, we do not consider fermionic sector. The obvious 
difference is that in \cite{rad2} they obtained type II action in Green-Schwarz formalism, while we study here pure spinor action. In \cite{rad3} a geometry of superspace is developed with type II model 
as the main example. One of the things discussed is relation of some sectors with pure spinor fields. At the end of the discussion it is useful to mention also the paper \cite{rad4}. In this article reduction 
of type IIA/B superstring theory from $10d$ to $9d$ is done, which is effectively could be T-dualization. The $L_\infty$ isomorphisms relate two coefficient $L_\infty$ algebras. Also they derive the 
Buscher rules for $RR$ field strength, which is done in this article using simpler mathematical methods.



\end{document}